\documentstyle[twoside, epsfig]{article}

\catcode`\@=11
\long\def\@makefntext#1{
\protect\noindent \hbox to 3.2pt {\hskip-.9pt
$^{{\eightrm\@thefnmark}}$\hfil}#1\hfill}               

\def\@makefnmark{\hbox to 0pt{$^{\@thefnmark}$\hss}}    

\def\ps@myheadings{\let\@mkboth\@gobbletwo
\def\@oddhead{\hbox{}
\rightmark\hfil\eightrm\thepage}
\def\@oddfoot{}\def\@evenhead{\eightrm\thepage\hfil
\leftmark\hbox{}}\def\@evenfoot{}
\def\sectionmark##1{}\def\subsectionmark##1{}}

\oddsidemargin=\evensidemargin
\newcounter{sectionc}\newcounter{subsectionc}\newcounter{subsubsectionc}
\renewcommand{\section}[1] {\vspace{12pt}\addtocounter{sectionc}{1}
\setcounter{subsectionc}{0}\setcounter{subsubsectionc}{0}\noindent
        {\tenbf\thesectionc. #1}\par\vspace{5pt}}
\renewcommand{\subsection}[1] {\vspace{12pt}\addtocounter{subsectionc}{1}
      \setcounter{subsubsectionc}{0}\noindent
      {\bf\thesectionc.\thesubsectionc.{\kern1pt \bfit #1}}\par\vspace{5pt}}
\renewcommand{\subsubsection}[1]
      {\vspace{12pt}\addtocounter{subsubsectionc}{1}
      \noindent{\tenrm\thesectionc.\thesubsectionc.\thesubsubsectionc.
      {\kern1pt \tenit #1}}\par\vspace{5pt}}
\newcommand{\nonumsection}[1] {\vspace{12pt}\noindent{\tenbf #1}
        \par\vspace{5pt}}

\newcounter{appendixc}
\newcounter{subappendixc}[appendixc]
\newcounter{subsubappendixc}[subappendixc]
\renewcommand{\thesubappendixc}{\Alph{appendixc}.\arabic{subappendixc}}
\renewcommand{\thesubsubappendixc}
        {\Alph{appendixc}.\arabic{subappendixc}.\arabic{subsubappendixc}}

\renewcommand{\appendix}[1] {\vspace{12pt}
        \refstepcounter{appendixc}
        \setcounter{figure}{0}
        \setcounter{table}{0}
        \setcounter{lemma}{0}
        \setcounter{theorem}{0}
        \setcounter{corollary}{0}
        \setcounter{definition}{0}
        \setcounter{equation}{0}
        \renewcommand{\thefigure}{\Alph{appendixc}.\arabic{figure}}
        \renewcommand{\thetable}{\Alph{appendixc}.\arabic{table}}
        \renewcommand{\theappendixc}{\Alph{appendixc}}
        \renewcommand{\thelemma}{\Alph{appendixc}.\arabic{lemma}}
        \renewcommand{\thetheorem}{\Alph{appendixc}.\arabic{theorem}}
        \renewcommand{\thedefinition}{\Alph{appendixc}.\arabic{definition}}
        \renewcommand{\thecorollary}{\Alph{appendixc}.\arabic{corollary}}
        \renewcommand{\theequation}{\Alph{appendixc}.\arabic{equation}}
        \noindent{\tenbf Appendix \theappendixc #1}\par\vspace{5pt}}
\newcommand{\subappendix}[1] {\vspace{12pt}
        \refstepcounter{subappendixc}
        \noindent{\bf Appendix \thesubappendixc. {\kern1pt \bfit #1}}
        \par\vspace{5pt}}
\newcommand{\subsubappendix}[1] {\vspace{12pt}
        \refstepcounter{subsubappendixc}
        \noindent{\rm Appendix \thesubsubappendixc. {\kern1pt \tenit #1}}
        \par\vspace{5pt}}

\newcommand{\smalllineskip}{\baselineskip=10pt}

\def\eightcirc{
\begin{picture}(0,0)
\put(4.4,1.8){\circle{6.5}}
\end{picture}}
\def\eightcopyright{\eightcirc\kern2.7pt\hbox{\eightrm c}}

\def\abstracts#1#2#3{{
        \centering{\begin{minipage}{4.5in}\baselineskip=10pt\footnotesize
        \parindent=0pt #1\par
        \parindent=15pt #2\par
        \parindent=15pt #3
        \end{minipage}}\par}}

\renewenvironment{thebibliography}[1]
        {\frenchspacing
         \ninerm\baselineskip=11pt
         \begin{list}{\arabic{enumi}.}
        {\usecounter{enumi}\setlength{\parsep}{0pt}
         \setlength{\leftmargin 12.7pt}{\rightmargin 0pt} 
         \setlength{\itemsep}{0pt} \settowidth
        {\labelwidth}{#1.}\sloppy}}{\end{list}}

\newcounter{itemlistc}
\newcounter{romanlistc}
\newcounter{alphlistc}
\newcounter{arabiclistc}

\newcommand{\fcaption}[1]{
        \refstepcounter{figure}
        \setbox\@tempboxa = \hbox{\footnotesize Fig.~\thefigure. #1}
        \ifdim \wd\@tempboxa > 5in
           {\begin{center}
        \parbox{5in}{\footnotesize\smalllineskip Fig.~\thefigure. #1}
            \end{center}}
        \else
             {\begin{center}
             {\footnotesize Fig.~\thefigure. #1}
              \end{center}}
        \fi}

\newcommand{\tcaption}[1]{
        \refstepcounter{table}
        \setbox\@tempboxa = \hbox{\footnotesize Table~\thetable. #1}
        \ifdim \wd\@tempboxa > 5in
           {\begin{center}
        \parbox{5in}{\footnotesize\smalllineskip Table~\thetable. #1}
            \end{center}}
        \else
             {\begin{center}
             {\footnotesize Table~\thetable. #1}
              \end{center}}
        \fi}

\def\@citex[#1]#2{\if@filesw\immediate\write\@auxout
        {\string\citation{#2}}\fi
\def\@citea{}\@cite{\@for\@citeb:=#2\do
        {\@citea\def\@citea{,}\@ifundefined
        {b@\@citeb}{{\bf ?}\@warning
        {Citation `\@citeb' on page \thepage \space undefined}}
        {\csname b@\@citeb\endcsname}}}{#1}}

\newif\if@cghi
\def\cite{\@cghitrue\@ifnextchar [{\@tempswatrue
        \@citex}{\@tempswafalse\@citex[]}}
\def\citelow{\@cghifalse\@ifnextchar [{\@tempswatrue
        \@citex}{\@tempswafalse\@citex[]}}
\def\@cite#1#2{{$\null^{#1}$\if@tempswa\typeout
        {IJCGA warning: optional citation argument
        ignored: `#2'} \fi}}

\def\@refcitex[#1]#2{\if@filesw\immediate\write\@auxout
        {\string\citation{#2}}\fi
\def\@citea{}\@refcite{\@for\@citeb:=#2\do
        {\@citea\def\@citea{, }\@ifundefined
        {b@\@citeb}{{\bf ?}\@warning
        {Citation `\@citeb' on page \thepage \space undefined}}
        \hbox{\csname b@\@citeb\endcsname}}}{#1}}

\def\@refcite#1#2{{#1\if@tempswa\typeout
        {IJCGA warning: optional citation argument
        ignored: `#2'} \fi}}

\def\refcite{\@ifnextchar[{\@tempswatrue
        \@refcitex}{\@tempswafalse\@refcitex[]}}

\def\pmb#1{\setbox0=\hbox{#1}
        \kern-.025em\copy0\kern-\wd0
        \kern.05em\copy0\kern-\wd0
        \kern-.025em\raise.0433em\box0}

\def\fnm#1{$^{\mbox{\scriptsize #1}}$}
\def\fnt#1#2{\footnotetext{\kern-.3em
        {$^{\mbox{\scriptsize #1}}$}{#2}}}

\def\fpage#1{\begingroup
\voffset=.3in
\thispagestyle{empty}\begin{table}[b]\centerline{\footnotesize #1}
        \end{table}\endgroup}

\def\runninghead#1#2{\pagestyle{myheadings}
\markboth{{\protect\footnotesize\it{\quad #1}}\hfill}
{\hfill{\protect\footnotesize\it{#2\quad}}}}
\font\tenrm=cmr10
\font\tenit=cmti10
\font\tenbf=cmbx10
\font\bfit=cmbxti10 at 10pt
\font\ninerm=cmr9

\font\eightrm=cmr8

\def\qed{\hbox{${\vcenter{\vbox{                      
   \hrule height 0.4pt\hbox{\vrule width 0.4pt height 6pt
   \kern5pt\vrule width 0.4pt}\hrule height 0.4pt}}}$}}

\newcommand{\beg}{\begin{equation}\label}
\newcommand{\enq}{\end{equation}}
\newcommand{\p}{\partial}

\newcommand{\bib}{\bibitem}
\newcommand{\nabl}{{\bf \nabla}}

\begin{document}

\runninghead {Vladimir V. Kassandrov}
{Singular sources of Maxwell fields $\ldots$}

\thispagestyle{empty}\setcounter{page}{1}
\vspace*{0.88truein}
\fpage{1}

\centerline{\bf SINGULAR SOURCES OF MAXWELL FIELDS}
\vspace*{0.035truein}
\centerline{\bf WITH SELF-QUANTIZED ELECTRIC CHARGE}
\vspace*{0.035truein}

\vspace*{0.37truein}
\centerline{\footnotesize Vladimir V. Kassandrov}

\centerline{\footnotesize \it
Department of General Physics, Russian People's Friendship University}
\baselineskip=10pt
\centerline{\footnotesize \it
Ordjonikidze Str. 3, 117419, Moscow, Russia,
e-mail: vkassan@sci.pfu.edu.ru}


\baselineskip 5mm

\vspace*{0.21truein}

\abstracts{Single- and multi-valued solutions of homogeneous Maxwell equations 
in vacuum are considered, with ''sources'' formed by the 
(point- or string-like) singularities of the field strengths and, generally, 
irreducible to any $\delta$-functions' 
distribution. Maxwell equations themselves are treated as consequences (say, 
integrability conditions) of a primary ``superpotential''field subject to 
some nonlinear and over-determined constraints (related, in particular, to 
twistor structures).  As the  
result, we obtain (in explicit or implicit algebraic form) a distinguished 
class of Maxwell fields, with singular sources necessarily carrying a 
``self-quantized'' electric charge integer multiple 
to a minimal ``elementary'' one. Particle-like singular objects are subject 
to the dynamics consistent with homogeneous Maxwell equations and undergo 
transmutations -- bifurcations of different types. The presented scheme 
originates from the ``algebrodynamical'' approach developed by the author 
and reviewed in the last section. 
Incidentally, fundamental equivalence relations between the solutions of 
Maxwell equations, complex self-dual conditions and of Weyl ``neutrino'' 
equations are established, and the problem of magnetic monopole is briefly 
discussed.}{}{}  


\bigskip

$$$$

\section{Introduction. Classical electrodynamics: problems and approaches}

It is generally believed that classical electrodynamics (CED) in vacuum is 
one of the most trustworthy and fundamental physical theories. However, it 
suffers from internal inconsistencies and paradoxes, some of them known  
over a century, yet not resolved so far. 

In CED we deal with two completely independent entities: (point-like) 
electric charges-sources and electromagnetic (EM) fields produced by and acting 
on them. As for mathematical structure, CED consists of Maxwell equations (ME) 
for fields and of equations of motion of sources 
under the action of the Lorentz force and the Abraham radiation-reaction 
force. This is a very complicated system of 
partial and ordinary differential equations. There is a lot of problems 
bearing on the system among which those of separation of external field from 
the proper field of the source [1] and of 
fitting the exact balance of the 
particles-fields' energy-momentum (for, say, a number of interacting 
point charges) seem to be crucial yet insoluble (in remind, e.g., of the 
paradox of {\it self-accelerating} solution). 

In hope to resolve these and other problems of CED and to describe some 
properties related to particles' structure (e.g. to explain the mystery of the 
quantized electric charge or to remove the divergence of the EM self-energy) 
a lot of {\it nonlinear} generalizations of Maxwell electrodynamics have been 
proposed [2] some of them being completely geometrical in origin. 

In nonlinear ED schemes field equations turn  
into linear ME asymptotically, in the limit of weak fields. On the other 
hand, they possess ``soliton-like'' solutions with {\it finite energy}, and 
equations of motion of such particle-like formations (under the assumption 
of stability) follow from nonlinear field equations themselves and,  
in the first approximation~\fnm{a}\fnt{a}{With respect to the small parameters -- 
the ratios $v/c$ and $R/R_0$, $v$ being relative velocity and $R$ -- separation 
of ``solitons'' while $R_0$ --- their radii},  lead to the Lorentz 
force~[3]. 

However, despite of a variety of new ideas and powerful methods developed, the 
project of nonlinear ED failed in the same way as it was with the 
program of geometrization of electromagnetism. Namely, no consistent and invariant 
equations of motion of particle-like field formations have been obtained, 
the origin of electric charge quantization (and all the more -- of the other 
quantum numbers of topological origin~[4]) hasn't been understood etc. 
Beyond any doubt, the principal difficulty here is that 
we are not at all aware of {\it how} to generalize CED and ME in particular, 
{\it which} nonlinearity and {\it which} it underlying geometry really {
\it encodes the Nature}. These questions will be briefly discussed below 
(section 7) in the framework of the so called {\it algebrodynamical} 
approach developed by the author.   

On the other hand, it turns out that linear ME themselves possess several 
classes of peculiar solutions [5-8], the ``knotted'' solutions with nontrivial 
topology  of field lines [5] and the solutions [7,8] with extended 
string-like or membrane-like singularities [7,8] among them. It is especially 
interesting that both types of solutions {\it necessarily carry self-quantized  
electric charge}, i.e. the charge integer multiple in value to some 
minimal {\it elementary} one [9, 10-12]. These solutions 
of ME can be naturally {\it selected} (and considered as {\it the only physically 
meaningful}) via some generating procedures in which ME are the direct 
consequences (say, the integrability conditions) of a highly nonlinear and/or 
{\it over-determined} system of field equations for a fundamental (scalar or 
spinor) primodial field. 

Thus, we meet there, in J.~Wheeler's terminology, with ``nonlinearity without 
nonlinearity'' or, in other words, with induced  or ``hidden'' nonlinearity 
[5,14]. This is a principally new paradigm alternative to that of the nonlinear 
ED. In its framework, ME completely preserve their linear form whereas all the 
restrictions on the shape and on ``quantum numbers'' of singular particle-like 
objects (as well as on their nontrivial time evolution) 
follow from the constraints imposed onto the primary generating field. 
In the {\it topological theory of electromagnetism} (TEE)
developed by A.F.~Ran\~ada and J.L.~Trueba [5,9,14] these constraints 
are specially fitted in the way to ensure ME to hold ``on shell'' and result in 
a complicated system of PDEs for generating scalars. On the other hand, in 
the {\it algebrodynamics} (AD) [10-13,15,7,8,21,29] these 
constraints originate from the {\it generalized Cauchy-Riemann equations} for 
biquaternion - valued functions - fields and are deeply related to {\it 
twistors} and to exceptional Weyl-Cartan geometries [11,21]. Consequently, 
the generating procedure can be reduced to resolution of a purely algebraic 
system of equations and opens thus the way to obtain the solutions of ME 
with extremely complicated structure of the ``particle-like'' 
(i.e. spacially bounded) singularities.   
 
The TEE and the AD approaches offer a quite new treatment of the 
``sources-fields'' problem. Whereas in CED we deal with {\it EM fields 
generated by charges} moving along their world lines, here we 
are brought to consider {\it the sources defined by the 
EM field itself in its singular locus}. By this, field 
singularities can be point-like~\fnm{b}\fnt{b}{In our terminology, the 
Coulomb field presents an example of a solution of the {\it source-free (!)} 
ME with a simplest point-like ``topological defect''} or extended 
and can consist of (a great number of) connected components (bounded or 
infinite in space).  On the other hand, 
EM field itself can be globally {\it multi-valued}, physically important example   
being presented by the EM field of the {\it Kerr-Newman solution} in GTR or by 
its flat analogue (see below, section 6). Singular ``sources''
of multi-valued solutions {\it in principle can not be described by any set 
of $\delta$-functions} so that the generally accepted in field theories and 
in mathematical physics 
paradigm of {\it distributions} seems to be not always sufficient and,  
moreover, not at all necessary and general (for this, see the papers [16-17]
and our discussion in [8,37]). It is especially important 
that singular sources always manifest themselves a well-defined and {\it finite} 
space and time distribution, and one has, in principle, no problem of divergence 
here, the problem which is insoluble in the framework of the accepted approach. 
                                                          
Historically, the concept of ``intrinsic'' singular sources of physical 
fields has been advocated, say, by H.~Bateman [18]  
et al. in the early XX century but unmeritely abandoned then in favour of the 
Dirac's $\delta$-function formalism. 

With respect to the above considerations and much similar to those used in 
the nonlinear ED paradigm, throughout the paper we 
deal with solutions of ``free'' linear ME (and also of other field equations) in 
the sense that {\it extended regular sources are absent} whereas the singular 
(generally extended) sources {\it of zero measure} are assumed to exist. 
To prevent any conflict with commonly used terminology, we shall avoid  
to call the considered ME ``free'' (and, all the more, ``source-free'') and 
will use the term ``homogeneous'' ME or simply ME themselves. The  
solutions are assumed to be analytical everywhere except in singular loci so 
that it's quite natural to consider their {\it complexification} which 
has been intensively advocated by E.T. Newman [19], R.~Penrose [27,30] et al. 
and which turns out to exhibit the self-dual structure of EM fields and 
to open the way to the solution of the charge quantization problem.
 
The main goal of the paper is to describe some simple constructions which 
relate the whole of the solutions of homogeneous ME to those of the complex 
self-dual conditions, of Weyl ``neutrino'' equations and of d'Alembert wave 
equation and which make it possible to generate the above solutions in a simple, 
effective and, in particular, in a purely {\it algebraical} way. In section 2 
we consider the complex and the self-dual structures of EM field. 
The proof of the charge quantization theorem is presented in section 3 where 
also the related problem of the {\it magnetic monopole} is briefly discussed. 
A simple generating construction for the whole class of (almost everywhere) 
analytical solutions of Maxwell and Weyl equations from one complex 
``superpotential'' function subject to wave equation 
is described in section 4 where we also establish important equivalence 
relations between the solutions of these fundamental equations. 
Further, in section 5, we present our main algebraic construction 
based on {\it twistor}  
structures and on the so called {\it Kerr theorem} and leading to Maxwell fields  
with self-quantized charges. Some examples of this class of solutions to ME are  
examined in section 6 where we are  mainly interested in the structure of field 
singularities. To conclude, in section 7 we present a review of the 
self-consistent algebrodynamical field theory and of its links with complex 
quaternionic analysis and with twistor structures and Kerr theorem.
  
To simplify the presentation, we do not use the differential forms or 
the 2-spinor formalism (for this, we refer the reader e.g. to our preprint [21]) 
and apply where possible the 3-vector notation.

\bigskip
\section{Self-dual complex nature of Maxwell fields}
\label{selfdual}

We consider the simplest system of homogeneous ME (the light 
velocity $c$ is taken to be $c=1$)
\begin{eqnarray}
\p_t ~{\bf H} = - \nabl \times {\bf E} , ~~~~\nabl \cdot {\bf H} = 0 ,\label{max1}\\
\p_t ~{\bf E} = + \nabl \times {\bf H} , ~~~~\nabl \cdot {\bf E} = 0 ,\label{max2}
\end{eqnarray}
where $\bf E$ and $\bf H$ are respectively the electric and the magnetic 
field strengths. First pair (\ref{max1}) of ME is identically satisfied 
by the (locally always existing) potentials $\phi(x),{\bf A(x)}$ such that 
\beg{pot1}
{\bf E} = -\p_t {\bf A} - \nabl \phi, ~~~~{\bf H} = \nabl \times {\bf A} .
\enq

Similarly, the second pair (\ref{max2}) can be identically satisfied by 
the (locally always existing) {\it conjugated} potentials $\psi(x),{\bf B(x})$ 
such that
\beg{pot2}
{\bf H} = -\p_t {\bf B} -\nabl \psi, ~~~~{\bf E} = - \nabl \times {\bf B} .
\enq

For consistency of definitions of field strengths the following constraints 
for potentials should hold:
\beg{constr}
\p_t {\bf A} + \nabl \phi = \nabl \times {\bf B}, ~~~~\p_t {\bf B} +\nabl \psi
= - \nabl \times {\bf A}.
\enq

On the other hand, for about a century it's well known (see e.g. [18]) 
that ME (\ref{max1}),(\ref{max2}) can be rewritten in a unified complex number form 
\beg{maxC}
i \p_t {\bf P}  =  \nabl \times {\bf P}, ~~~~\nabl \cdot {\bf P} = 0 ,
\enq
say, for complex vector ${\bf P} = {\bf E} + i {\bf H}$. Consistency conditions 
(\ref{constr}) also admit a complex representation of the form 
\beg{sd}
\p_t {\bf C} + \nabl \Pi + i \nabl \times {\bf C} = 0 
\enq
with complex potentials $\Pi = \phi + i \psi$ and ${\bf C} = {\bf A} + i{\bf B}$ via 
which the complex field strength vector ${\bf P}$ is defined as follows:
\beg{potC}
{\bf P} = -\p_t {\bf C} -\nabl \Pi =  i \nabl \times {\bf C} 
\enq
and, provided the consistency conditions (\ref{sd}) are fulfilled, 
turns to identity the complex-form ME (\ref{maxC}). 

Thus, locally ME reduce to and are equivalent to three {\it first-order (!)} 
equations (\ref{sd}) for complex 4-potential vector $\Pi(x),\bf C(x)$. 
They are evidently gauge invariant (with gauge parameter being arbitrary 
(smooth) {\it complex} function of coordinates). Therefore, one can 
complete the system (\ref{sd}) by a gauge fixing condition, e.g. by 
the complex Lorentz gauge
\beg{lorgauge}
\p_t \Pi + \nabl \cdot {\bf C} = 0 .
\enq

On the other hand, Eqs.(\ref{sd}) admit alternative treatment. Let us 
from the very beginning consider the {\it complex-valued} EM fields $\vec 
{\cal E}, \vec {\cal H}$  defined via holomorphic 4-potentials $\Pi(x),
{\bf C(x)}$ in a usual way,
\beg{strC}
\vec {\cal E} = -\p_t {\bf C} - \nabl \Pi , ~~~~\vec{\cal H} = \nabl \times {\bf C}. 
\enq      

Then Eqs.(\ref{sd}) turn to be just the {\it antiself-duality 
conditions} for complex field strengths $\vec{\cal E}, \vec{\cal H}$, 
\beg{sdeh}
\vec {\cal E} - i \vec {\cal H} \equiv -\p_t {\bf C} - \nabl \Pi - i \nabl \times {\bf C} = 0 , 
\enq
and, if desirable, Eq.(\ref{lorgauge}) can be joined to the latters. Note that 
the complex conjugated fields turn then to be {\it self-dual}. Eqs.(\ref{sdeh}) 
(and also those for conjugated fields) below will be called {\it complex self-duality} 
(CSD) conditions [10,11]. Complex self-dual EM fields have been considered, e.g., 
by R.~Penrose [27] as the ``wave functions of photons''. G.~A.~Alekseev [22] 
used the CSD conditions instead of Maxwell equations in his seach of the 
solutions of the Einstein-Maxwell and related equations. The Heyl-Obukhov 
``metric-free ED'' [23] can also be thought of as a nonlinear extension  
of the CSD conditions.  

As to the number of degrees of freedom, for CSD fields it is the same as for real 
Maxwell fields: the procedure of complexification redoubles the number but 
the SD conditions reduce it to the ordinary one. Indeed, let $\vec {\cal E} = 
{\bf E} + i {\bf D}, ~\vec {\cal H} = {\bf H} + i {\bf N}$ where ${\bf E, D, 
H, N}$ are the real and imaginary parts of complex electric and 
magnetic field strengths respectively. Then CSD conditions (\ref{sdeh}) 
result in 
\beg{algconst}
{\bf N} = - {\bf E}, ~~~{\bf D} =  {\bf H} 
\enq
so that only, say, the real-part fields remain (algebraically) independent while 
the imaginary-part ones are {\it dual} to them.

Dynamically, every solution $\Pi(x),{\bf C(x)}$ of CSD conditions (\ref{sdeh}) 
corresponds to a solution of {\it complexified} ME and, by virtue  
of {\it linearity} of the latters, -- to a {\it pair of solutions} of real 
ME for the set of field strengths ${\bf E},{\bf H}$ and for its dual one. For 
example, one gets
$$
\nabl \cdot \vec {\cal H} = \nabl \cdot (\nabl\times {\bf C}) \equiv 0, ~~
\Rightarrow ~~ \nabl \cdot \vec {\cal E} = i \nabl \cdot \vec {\cal H}=0 ~~~ 
and ~so ~on.
$$
The converse statement has been already proved: for every solution to 
homogeneous ME 
locally some complex potentials can be defined subject to CSD conditions.

Thus, we can declare that {\it locally ME are completely equivalent to the 1-st 
order CSD equations}. It can be easily checked also that the ordinary 2-d order 
d'Alembert equations are just the integrability conditions of CSD system  
(\ref{sdeh}). Therefore, all EM fields (contrary to the well-known case of 
nonAbelian Yang-Mills fields) can be regarded as complex self-dual in nature.

It's interesting to ask for the reasons to select the $\Re$-part 
(or its dual $\Im$-part) fields from the primodial {\it holomolphic} 
field strengths $\vec{\cal E},\vec{\cal H}$. This may be related to a 
peculiar fact that {\it for complex Maxwell fields all of the components of 
the energy-momemtum tensor are identically zero.} In particular, for 
the energy density $W$ on account of CSD conditions (\ref{sdeh}) one gets  
$$
W \propto (\vec {\cal E}^2 + \vec {\cal H}^2) \equiv 0 ~~~ and ~so ~on.
$$
One can say in jest that {\it complex fields do not possess any energy and 
acquire it only through self-division into real and imaginary parts} via 
which the conservation laws can be constructed in a usual way. 

The CSD structure of EM fields which consolidates them with the other well-known 
quantum  fields is, in our opinion, not a formal 
renotation but a fundamental property which manifests itself even in 
the complex structure (\ref{maxC}) of homogeneous ME themselves~\fnm{c}\fnt{c}{Owing to 
different signs in the r.h.s. of the ``curl'' Eqs. (\ref{max1}),(\ref{max2}) one 
fails to represent the ME in, say, the {\it double-number} algebraic form (for  
which $i^2 = 1$) but only via the exeptional algebra of complex numbers with 
$i^2=-1$}. 

Moreover, we can conjecture that the CSD conditions, though being locally 
equivalent to homogeneous ME,
are more fundamental from general viewpoint. Indeed, we demonstrate 
below (section 3) that exchange of ME by CSD makes it possible to propose a 
peculiar solution of the charge quantization problem. Besides, it restores 
the complete intrinsic electric-magnetic symmetry of ME and  
offers a new approach to the magnetic 
monopole problem, another than that proposed by P.A.M.~Dirac. Finally,  
establishment of (local) equivalence of ME and CSD conditions opens the 
way to powerful ``superpotential'' procedures for generation of  
complicated solutions to both systems (sections 4,5) and, 
somewhat mysteriously, permits to relate these solutions with those of the Weyl  
neutrino equation up to establishment of the full local equivalence of these 
equations (section 4).

\bigskip
\section{Quantization of electric charge and electric-magnetic symmetry in 
the complex self-dual electrodynamics}

Quantization of electric charge follows from CSD conditions if a   
gauge invariant interaction with another field $\Psi(x)$ 
(a ``section'' of a scalar, 2-spinor, bispinor etc. fibre bundle) is included, so 
that {\it complex-valued} EM 4-potentials $C_\mu(x)$ enter the theory only 
through the form of the ``lengthened'' derivative $(\p_\mu - bC_\mu)\Psi$, 
~$b=const$ being the coupling constant. The theory is, therefore, invariant 
under the gauge transformations 
\beg{gtrans}
\Psi \mapsto e^{\alpha(x)} \Psi, ~~~~C_\mu \mapsto C_\mu + b^{-1} \p_\mu \alpha 
\enq
where $\alpha(x)$ is any smooth {\it complex} function of coordinate. 

Apart from equations for the field $\Psi(x)$ and {\it instead of 
inhomogeneous ME} (with {\it distributed} sources generated by the $\Psi$-induced 
charge-current density $j_\mu(x)$) we assume here the CSD conditions (\ref{sdeh}) 
for the complex-valued EM fields to be satisfied. A particular example of such a theory 
will be presented in section 5. Consider now the solutions of this system for which the 
field strength singularities -- point-like or extended charge carriers -- are 
{\it localized in a bounded region of 3-space} (at any finite moment of time). 
We shall call them {\it singular particle-like} (SPL) solutions below. Then we  
formulate the following 

{\bf Theorem.} For every SPL solution for which the function $\Psi(x)$ is 
single-valued everywhere except the singularities of field strengths {\it the 
value of electric charge is either zero or integer-multiple of 
a minimal "elementary charge" (equal to $q_{min} = 1/(2b)$)}.

The proof of this theorem [12] exploits, besides the CSD conditions 
themselves, also the well-known Dirac's considerations on the magnetic monopole 
problem [24], namely the ordinary quantum field conjectures on the {\it existence} 
of EM potentials as of essential physical quantities (i.e. not only of the field 
strengths themselves) 
and on the {\it gauge invariance} (\ref{gtrans}) of the system of field equations 
under study. Making use of the definition of electric charge $q$ via the {\it 
Gauss theorem} and integrating over the closed 2-surface $\Sigma$ {\it enveloping 
one or more connected and bounded singularities} we have on account of the CSD 
conditions
\beg{gauss}
4\pi q = \oint_\Sigma \vec {\cal E} d{\vec \sigma} = i\oint_\Sigma 
\vec {\cal H} d{\vec \sigma} \equiv 4\pi i\mu ,
\enq
so that the electric charge $q=i\mu$ where $\mu$ is the correspondent {\it magnetic 
charge} of the enclosed singularity (singularities).                       
Thus, in theories of the type considered we deal in fact with {\it dions}                                  
carrying {\it always equal in modulus electric and magnetic charges.} We 
shall return to discuss the magnetic monopole problem at the end of this 
section.
 
Let us assume now that complex vector-potential $\bf C(x)$ is analytic 
everywhere except on a set of one-dimensional subspaces -- ``Dirac's strings'' -- 
where it turns to infinity~\fnm{d}\fnt{d}{In principle, singular locus of   
potential $\bf C(x)$ can also contain {\it two-dimensional} components, 
an example of this situation being studied in our works [13,21].  
However, it's easy to see that these singular subspaces do not contribute to 
the charges' integrals (\ref{gauss})}. 
Such singular strings inevitably exist for 
nonzero $\mu$ since, otherwise, the field is solenoidal everywhere, $\vec 
{\cal H} = \nabl \times \bf C$, and the magnetic flow in (\ref{gauss}) is null. 
Taking this into account, we can cut from the surface $\Sigma$ the whole 
set of {\it infinitesimal disks} \{D\} which are pierced by the strings 
and can pass in this way to integrate in Eq.(\ref{gauss}) over (now not closed) 
2-surface $\Sigma^\prime=\Sigma-\{D\}$ across which the potential $\bf C$ is 
now regular. Notice that this procedure doesn't change the value of 
flows (\ref{gauss}) since the field strengths themselves are regular 
{\it everywhere on $\Sigma$ including the disks $\{D\}$} so that the flows 
through the disks are surely infinitesimally small.

Now, making use of the {\it Stokes theorem} for the surface $\Sigma^\prime$, 
we can reduce expression (\ref{gauss}) to a sum of integrals  over its  
{\it boundary}, i.e. over a set of infinitesimal closed loops $L\in \Sigma$ 
encircling the ``strings'', 
\beg{curve}
4\pi \mu = \sum \oint_L {\bf C}d{\bf l} .
\enq
Only {\it singular} part of potential contributes into the latters integrals. 
This part is  necessarily a {\it pure gauge}, ~ ${\bf C}_{sing} = \nabl \Lambda$ , ~
$\Lambda(x)$ being a complex function. Indeed, if it' were not the case 
the magnetic field would be also singular on $L\in \Sigma$ what contradicts to the 
assumptions of the theorem. Therefore, every integral in (\ref{curve}) 
reduces to $\Delta \Lambda$, i. e. to the increment of the (multi-valued) 
function $\Lambda(x)$ in bypassing the closed loop. Now, taking in account the 
gauge invariance of the theory (\ref{gtrans}) and the {\it single-valuedness}  
of the field $\Psi(x)$ outside of the field strengths' singularities we get 
\beg{sval}
\Delta \Lambda = b^{-1} \ln \Delta \Psi = b^{-1} 2\pi i n , ~~~n \in Z ,
\enq
and from Eqs.(\ref{gauss}) and (\ref{curve}) obtain finally for 
electric $q$ and magnetic $\mu$ charge of a singular source 
\beg{quanta}
q = i\mu = \frac{N}{2b}, ~~~   N=\sum n = 0,\pm 1, \pm 2, ...  .
\enq
                                                               
To conclude, let us compare our results with Dirac's construction. With 
respect to quantum theory, he took $b = ie / \hbar c$ and came from here 
to well-known {\it quantization constraint} [24]
\beg{dirquant}
\mu e = (\hbar c/2)N 
\enq
as it follows from our formula (\ref{quanta}) too. However, the Dirac  
condition (\ref{dirquant}) doesn't fix the values of each of the charges 
but only relates them one to another. Besides, the quantity $e$ therein 
plays the role of the coupling constant and, generally, has nothing to 
do with the charge of a particle-like (singular or regular) solution of 
field equations which therein remains indefinite. Contrary, in our construction  
the quantity $q$ has the explicit meaning of a characteristic 
of admissible field distributions. If we choose the coupling constant 
to be (in dimensional units) $b = 1/2e$ then the minimal (elementary) 
electric charge will be $q_{min} = e$. Below (sections 5,7) we demonstrate 
that in twistor and in the algebrodynamical theories such solutions really 
exist and in the simplest 
case describe a self-quantized Coulomb-like singularity carrying the charge 
necessarily {\it equal (i.e not multiple) to the elementary one}.  

Let us say also a few words about the magnetic charge in the framework of 
the above presented scheme. Its physical meaning and {\it reality} completely 
depends on the {\it dynamics} of singular sources, i.e. on the structure of 
the Lorentz and other forces acting on the dion, and it would be speculative 
to discuss it here in detail. 
Nonetheless, let us assume that, with respect to the quantization theorem, {\it 
only one sort of particles does exist, with necessarily equal in modulus 
electric and magnetic charges}. Then such a ``CED with magnetic charge''
(see e.g. [25]) is known to reduce (by {\it dual rotation} of 
electric - magnetic fields) to the scheme in which only one {\it effective} 
charge (which can be regarded, as an option, to be electric or magnetic one) 
manifests itself. Note that for complex fields dual transformation of 
$\Re$-parts $\vec {\bf E},\vec {\bf H}$ corresponds to multiplication 
by the phase factor $e^{i\theta}$ which obviously preserves the  
property for {\it complex} fields to be (anti)self-dual, 
$\vec {\cal E} - i\vec {\cal H}= 0$. 

We conjecture, therefore,  
that in gauge field theories with minimal interaction with (complex) EM field 
for which CSD conditions replace Maxwell equations {\it we deal only with singular 
sources carrying one effective (say, electric) charge}; from this viewpoint 
{\it magnetic monopoles do not manifest themselves at all}. Other geometrical 
considerations which partially support this conclusion can be found in [12].

\section{Generating constructions and equivalence between the solutions 
of Maxwell and Weyl equations}

Let us continue to disscuss the properties of CSD conditions (\ref{sdeh}) for 
complex 4-potentials $C^\mu(x)$ which are locally equivavent to homogeneous ME. 
It turns out that 
for {\it every} solution of the latters a gauge can be choosed in  
which only {\it two} complex components of 4-potential are dictinct from 
zero. Moreover, in this gauge CSD conditions reduce to well-known 
{\it Weyl equations} ``for neutrino'',  and {\it every} solution of them 
both (and, consequently, of ME themselves) 
can be generated via differentiation of a complex {\it one-component(!)} 
function subject to wave d'Alembert equation.

To prove these statements let us introduce the commonly used  
{\it spinor} or {\it null} space-time coordinates  $u = t + z , ~ v = t - z , 
~w = x - i y , ~\bar w = x + i y$ which can be collected to form the 
{\it Hermitian} $2\times 2$-matrix of coordinates
\beg{hermit}
X=X^+ = \left(
\begin{array}{cc}
u & w \\
\bar w & v 
\end{array}\right)
\enq
Similarly, for complex 4-vector of potentials $C^\mu(x)$ its matrix components 
look as follows: $C_u = \Pi + C_3, ~C_v = \Pi - C_3, ~C_w = C_1 - iC_2 ,  
~C_{\bar w} = C_1 + iC_2$ (note that $\Pi,~C_a=-C^a,~ a=1,2,3$ are  
the components of the {\it covariant} 4-vector $C_\mu$) and form a matrix of 
general type 
\beg{spinpot}
C = \left(
\begin{array}{cc}
C_u & C_w \\
C_{\bar w} & C_v 
\end{array}\right)
\enq

Now it is easy to make sure that CSD conditions (\ref{sdeh}) 
together with gauge condition (\ref{lorgauge}) can be 
rewritten in the following (splitting into two pairs of independent equations) 
form of {\it double - Weyl equations} (DWE):
\beg{dweyl}
\begin{array}{ccc}
\p_{\bar w} C_u = \p_u C_w & \p_v C_u = \p_w C_w  \\
                           &                        \\
\p_{\bar w} C_{\bar w} = \p_u C_v & \p_v C_{\bar w} = \p_w C_v ,
\end{array}
\enq
or, equivalently, in the following matrix form:
\beg{mweyl}
C \overleftarrow W   = 0,  ~~~W \equiv (\p_t - \vec {\bf \sigma} \cdot \nabl)
\enq
where $W$ is the Weyl differential operator (acting here to the left as 
indicated by the arrow) defined via three {\it Pauli matrices} $\vec {\bf \sigma}$ 
in the usual representation.

Thus, in the Lorentz gauge CSD equations are equivalent to taken twice the 
Weyl equations for the ``spin 1/2 particles with zero rest-mass''. By this, 
the complex 4-vector of potentials can 
be equivalently treated as {\it a pair of Weyl 2-spinors} $\psi^{(0)}= 
\{C_u, C_w\}$ and $\psi^{(1)}=\{C_{\bar w}, C_v\}$. The two alternative 
representations are possible owing to the wide group of invariance of DWE 
(\ref{mweyl}) with respect to Lorentz transformations 
\beg{lortrans}
X \mapsto A X A^+ , ~~~W \mapsto (A^+)^{-1} W A^{-1} ~~~  
C \mapsto D C A,
\enq
where $A$ and $D$ being arbitrary $SL(2,C)$-matrices. In particular, 
if we take $D=A^+$, the matrix $C$ will transform as a (covariant) complex 
4-vector whereas for $D = Id$ (the unity matrix) the two columns of $C$ will 
tranform independently as a pair of 2-spinors $\psi^{(A)}, ~A=0,1$.  

Dualistic vector-spinor nature of complex EM potentials follow 
thus from the fact that the fields subject to the DWE (\ref{dweyl}) 
transform by  
{\it reducible} representation of Lorentz group so that there is no 
contradiction with generally accepted point of view (see also [11]).

As the next step let us notice now that for all the solutions of DWE 
a $2\times 2$ {\it superpotential} matrix $M(x)$ which turn Eqs.(\ref{dweyl}) 
to identity does exist locally,  
\beg{superpot}
C =  M \overleftarrow W^* ,  ~~~~ M\overleftarrow W^* \overleftarrow W 
\equiv \Box M = 0,
\enq 
where the conjugated Weyl operator $W^* = \p_t + \vec{\bf\sigma} \cdot \nabl$  
and the 2-d order d'Alambert operator $\Box \equiv WW^* = \p_{tt} - \Delta$ 
are defined. Such a matrix $M(x)$ can be always found since 
the integrability conditions for Eq.(\ref{superpot}) 
are just the DWE for matrix $C(x)$ and, by assumption, are satisfied. 

Now we can easily restore the gauge invariance of the procedure neglecting 
for  this Lorentz gauge condition (\ref{lorgauge}) which 
corresponds to the {\it trace} part of matrix Eq.(\ref{dweyl}). Then CSD conditions 
(\ref{sdeh}) themselves are equivalent to the trace-free part of DWE.    
Then, together with the gauge freedom to choose the generating matrix 
\beg{Mgauge}
M \mapsto M + \Gamma \overleftarrow W, ~~~\Box \Gamma = 0,
\enq
($\Gamma(x)$ being any matrix with components subject to wave equation) 
with respect to (\ref{superpot}) {\it preserving} the potentials $C(x)$, we  
whould get again the gauge freedom of complex potentials of usual type,
\beg{Cgauge}
C_\mu \mapsto C_\mu + \p_\mu \alpha,
\enq 
($\alpha(x)$ being arbitrary complex function) as well as the {\it resudial} 
gauge invariance 
\beg{resgauge}
C_\mu \mapsto C_\mu + \p_\mu \lambda, ~~~\Box \lambda = 0,
\enq
($\lambda(x)$ being any complex function subject to wave equation). 
Both types of gauge transformations do not change the  
field strengths correspondent to potentials $C_\mu(x)$, and transformations 
(\ref{resgauge}) preserve also the Lorentz gauge condition (\ref{lorgauge}) 
being a symmetry of the whole system (\ref{dweyl}) of DWE~\fnm{e}\fnt{e}{The 
extended symmetry $C\mapsto C + \Lambda\overleftarrow W^*, ~~\Box \Lambda = 0$ 
with $\Lambda(x)$ being instead a full {\it matrix} also preserves the structure 
of DWE but {\it changes} field strengths and results in a different solution 
to complexified ME}. To conclude, we have proved that 

\noindent
{\it every solution of CSD conditions can be locally 
obtained via differentiation of a $2\times2$ complex matrix field subject 
to d'Alembert equation. Every solution of homogeneous ME can be obtained from 
here via second differentiation and subsequent separation of real and  
imaginary part of complex field strengths. The procedure is completely 
Lorentz and gauge invariant}
~\fnm{f}\fnt{f}{Under Lorentz transformations the superpotential matrix $M(x)$ 
can be found to behave as a pair of (conjugated) spinors}. 
                             
Further on we shall use the above-mentioned gauge freedom to simplify as much 
as possible the 
choice of complex potentials and superpotentials generating the solutions of ME. 
At first we fix the Lorentz gauge (\ref{lorgauge}) and come, therefore, to the 
structure of the full system of DWE (\ref{dweyl}). Then we exploit the residual 
gauge invariance (\ref{resgauge}) and choose the parameter $\lambda(x)$ in a 
way that some two components of the transformed potentials (constituting one of 
the spinors $\psi^{(A)}(x)$, say $\psi^{(0)} = \{C_u,C_w\}$) will turn to 
zero~\fnm{g}\fnt{g}{This is always possible since the integrability conditions 
for equations on desirable parameter $C_u = \p_u \lambda,~C_w =\p_{\bar w} 
\lambda$ hold 
identically on account of the first pair of Eqs.(\ref{dweyl}) themselves}. Thus, 
for every solution of ME we were able to reduce the DWE for potentials to 
ordinary Weyl equations (WE) for one 2-spinor, say to $\psi^{(1)} = 
\{C_{\bar w},C_v\}$ represented by two nonzero components of complex potentials.  
                
As the last step we notice that under the above choice of potentials the 
generating superpotential matrix $M(x)$ also reduce to one 2-spinor and, 
moreover, can be brought to only one nontrivial component by use of its 
own gauge freedom (\ref{Mgauge}). In our particular choice we satisfy  
WE represented by the first remaining pair of Eqs.(\ref{dweyl}) setting e.g. 
\beg{fingauge}
C_{\bar w} = \p_u G, ~~~C_v = \p_{\bar w} G, ~~~ \Box G = 0,
\enq
where $G(x)$ is a ``superpotential'' complex function subject to wave equation. 

At this point we can formulate our final result.

\noindent
{\it Every solution of Weyl equations can be obtained via differentiation from 
one-component complex function subject to wave equation. Every solution of 
homogeneous Maxwell equations can be obtained from here via repeated  
differentiation}. 

It should be emphasized that a remarkable equivalence relation of homogeneous 
Maxwell and Weyl equations has been also established: for {\it every} solution 
of ME locally a {\it two-component} complex potential can be defined which 
satisfies WE. Every 
solution of WE leads via differentiation to a solution of (complexified) ME.    
   
Many interesting and physically important questions do arise in the framework 
of the proved equivalence relation. For example, one can write out   
at least two {\it nonequivalent} expressions for ``energy'', 
``angular momentum'', Maxwell or Neuter's charge and other conserved quantities,  
for any pair of related solutions of Maxwell or Weyl equations, which follow 
from the distinct structure of Lagrangians for those fields. This question, 
as well as that about generalized continious and discrete 
symmetries of ME and WE, deserve special consideration.  
                                                 
Finally, let us present a simplest example of the Coulomb-like solution of ME which 
satisfies them everywhere except the singular point. For this, 
we can make, say, the following choice of the superpotential $G(x)$ and of 
the two components of complex potentials (\ref{fingauge}) respectively:
\beg{coulomb}
G=\frac{4q\bar w}{z+r}, ~~~~~C_{\bar w} = \p_u G = -\frac{2q\bar w}{z+r}, ~~~
C_v = \p_{\bar w} G = \frac{2q}{r} ,
\enq
where $z=(u-v)/2, ~~r=\sqrt{w\bar w + z^2} =\sqrt{x^2+y^2+z^2}$ and the 
constant $q$ is assumed to be real. 
The components $\psi=\{C_{\bar w}, C_v\}$  constitute then a spinor which 
satisfies WE everywhere {\it except the "open string"} $x=y=0,~z\le 0$ where  
it turns to infinity. On the other hand, these components correspond to 
the complex field strengths
\beg{cfs}
{\cal E}_a = i{\cal H}_a = \frac{qx_a}{r^3}, ~~~a=1,2,3
\enq
the real part of which represents the Coulomb electric field of the point 
charge $q$ while the imaginary part -- the dual field of the magnetic 
monopole with the charge $\mu=-q$ equal in modulus to the electric one.

\section{Twistor generating construction and Maxwell fields with 
self-quantized charge} 

In the preceding section a simple ``superpotential'' construction was  
described 
via which the whole of solutions of homogeneous ME can be obtained starting from 
those of complex d'Alembert equation.  This construction though interesting from 
technical point of view has little to do with the basic approach considered in 
the paper, i.e. with the paradigm of {\it induced nonlinearity} and with methods 
to select a 
subclass of physically interesting (effectively interacting and self-quantized) 
solutions of ME through imposing natural and strong restrictions on 
generating functions (on potentials or superpotentials) themselves. Realization 
of such a procedure needs the use of twistor structures and of the Kerr 
theorem to which we pass now. 

Let $\xi(x)$ and $\tau(x)$ be two 2-spinor fields for which the following 
linear {\it incidence relation} takes place~\fnm{h}\fnt{h}{For simplicity, we 
do not distinguish here between primed and unprimed spinor indices and 
neglect the usually settled multiplier "i" in the incidence relation below}:
\beg{inc}
\tau = X\xi ~~\Leftrightarrow ~~\tau^0 = u\xi_0 + w\xi_1, ~~\tau^1 =
\bar w \xi_0 + v \xi_1,
\enq
where $X=X^+$ is the Hermitian matrix (\ref{hermit}) of space-time coordinates 
represented by $u=t+z,~v=t-z,~w=x-iy,~\bar w=x+iy$.  
The pair of spinors $\{\xi(x),\tau(x)\}$ linked via incidence relation with the points of 
Minkowsky space-time forms the so called {\it null twistor} field [27]. 

Eq.(\ref{inc}) is evidently form-invariant under scaling of the both spinors 
together and, therefore, we can pass to three {\it projective} components of 
the twistor which form the complex projective space $CP^3$. Specifically, 
we set the component $\xi_0=1$ (assuming it to be nonzero in the region 
of space-time considered) and reduce the twistor to only three projective 
components  
\beg{twistcomp}
\xi_1 \equiv G, ~~~\tau_0 = u + wG, ~~~\tau_1=v+\bar w G.
\enq 

Let us demand now that {\it for some twistor field its three projective components 
are functionally dependent} as functions of space-time coordinates. In other 
words, let there exist a function $\Pi(G,\tau^0,\tau^1)$~\fnm{i}\fnt{i}{We consider 
$\Pi$ itself to be analytical with respect to its three complex arguments}
of three complex variables (\ref{twistcomp}) for which the equation
\beg{kerr}
\Pi(G,\tau^0,\tau^1) = \Pi(G,~wG+u,~vG+\bar w) = 0 
\enq
holds identically for every space-time point $X=\{u,w,\bar w,v\}$. Eq.(\ref
{kerr}) has been introduced firstly by R.P. Kerr [28] and is based 
on fundamental geometrical and algebraical structures deeply 
related to physical space-time and field dynamics (they will be briefly 
discussed at the end of the section). Note also that, formally, {\it Kerr 
functional condition} (KFC) (\ref{kerr}) defines a {\it hypersurface} in 
$CP^3$-space. 

Let us verify now that KFC itself restricts the admissible 2-spinor (twistor) 
fields and, in a remarkable way, gives rise to a physically significant class of 
the latters. Indeed, Eq.(\ref{kerr}) can be {\it algebraically} and 
{\it continiously} resolved with respect to the only unknown $G$ at every 
space-time point (except those at which $G(x)$ has poles or branching points, 
see below). In this way we come to an ``almost everywhere'' analytical and 
single-valued branch of (globally multi-valued) complex field defined at a 
region of space-time. 

Rather unexpectedly, this algebraically generated field $G(x)$ 
satisfies a whole number of fundamental Lorentz invariant differential 
equations, in particular the wave and the eikonal equations [26,7]. 
To prove this, let us differentiate the 
KFC (\ref{kerr}) with respect to coordinates $u,w,\bar w, v$ and get then 
with respect to  Eq.(\ref{twistcomp}) 
\begin{equation}\label{kerrder}
\begin{array}{cc}
\p_u G = - P^{-1} \Pi_0, & \p_w G = - P^{-1} G\Pi_0,  \\

\p_{\bar w} G = - P^{-1} \Pi_1, & \p_v G = - P^{-1} G\Pi_1 ,  
\end{array}
\enq
where $\Pi_C, ~C=0,1$ are the derivatives of $\Pi$ with respect to 
correspondent twistor arguments $\tau^C$ and $P = d\Pi / dG$ is the {\it total 
derivative} of $\Pi$ taken on account of the constraints (\ref{twistcomp}). In 
the {\it branching points} which are defined by the condition
\beg{sing}
P = \frac{d\Pi}{dG} = 0 
\enq
the derivatives of $G(x)$ become singular.  In regular region, eliminating the 
quantities $\Pi_C$ from Eq. (\ref{kerrder}) we get two 
nonlinear differential constraints for the derivatives of $G(x)$,
\beg{sfc}
\p_w  G  = G \p_u G , ~~~\p_v  G = G \p_{\bar w} G 
\enq
for which (as it is easy to prove, see e.g. [27]) KFC (\ref{kerr}) 
represents their {\it general solution}. As a direct consequence of 
(\ref{sfc}), multiplying the two equations we get the nonlinear eikonal equation 
\beg{eik}
\vert \underline{\nabla} G \vert^2  \equiv 4(\p_u G \p_v G- \p_w G \p_{\bar w} G) = 0,
\enq
and calculating the integrability conditions -- the linear wave equation
\beg{Gwave}
\Box G = 4(\p_u\p_v G - \p_w \p_{\bar w} G) = 0,
\enq
which should hold both together in consequence of KFC (\ref{kerr}) or 
correspondent differential constraints (\ref{sfc}). 

Since the complex field $G(x)$ satisfies wave equation, by virtue of the 
results of the previous section it can be taken as a {\it superpotential} 
function to generate the solutions of ME and WE (with singular sources). 
Fot this, we identify the two nonzero components of complex potentials with 
correspondent derivatives
\beg{Gpot}
C_{\bar w} = \p_u G = - P^{-1} \Pi_0,   ~~~C_v = \p_{\bar w} G = - P^{-1} \Pi_1 
\enq 
and verify immediately that they satisfy WE of the form identical to that  
represented by the second pair of Eqs.(\ref{dweyl})
\beg{Gweyl}
\p_{\bar w} C_{\bar w} =\p_u C_v, ~~~ \p_v C_{\bar w} = \p_w C_v 
\enq
by virtue of definitions (\ref{Gpot}) and of wave equation (\ref{Gwave}) 
respectively. On the other hand, WE (\ref{Gweyl}) can be considered as 
a reduced system of CSD conditions (\ref{sdeh}) complemented by the Lorentz 
gauge condition (\ref{lorgauge}). Therefore, the antiself-dual
($\vec {\cal E} = i\vec{\cal H})$ complex field 
strengths can be defined, of the form (${\cal E}_\pm \equiv {\cal E}_1 \pm i {\cal 
E}_2$)
\beg{Gstreng}
{\cal E}_3 =  -\p_u C_v = - \p_u\p_{\bar w} G , ~~  
{\cal E}_+ = \p_u C_{\bar w} = \p_u\p_u G , ~~
{\cal E}_- = -\p_{\bar w} C_v = -\p_{\bar w}\p_{\bar w} G  
\enq
which {\it satisfy homogeneous ME for every $G(x)$ 
implicitly defined via KFC (\ref{kerr})}. 

As well as the potentials (\ref{Gpot}), the field strengths (\ref{Gstreng})
can be expressed via the (1-st and 2-d order) derivatives 
$\Pi_C,~\Pi_{CD},~ C,D=0,1$ of generating function $\Pi$ with respect to 
its twistor arguments $\tau^C$. Final expression for (symmetric) {\it spinor 
of (antiself-dual) electromagnetic field} $F_{CD}=\{-{\cal E}_+,
{\cal E}_3,{\cal E}_-\}$ has been obtained in [21,29] 
and has the following invariant form:
\beg{emsp}
F_{CD} = \frac{1}{2P}\left\{\Pi_{CD} - 
\frac{d}{dG}\left(\frac{\Pi_C\Pi_D}{P}\right)\right\}.
\enq
Comparison of this Eq.(\ref{emsp}) with condition (\ref{sing}) demonstrates 
then that {\it singularities of field strengths (\ref{Gstreng}) occur  
just in the branching points of superpotential function $G(x)$} so 
that {\it everywhere except in the singular locus of EM field the principal 
field $G(x)$ is necessarily single-valued}.

In the considered twistor construction the {\it harmonic} superpotential function 
$G(x)$, $\Box G = 0$ is restricted to a much more extent being subject to 
nonlinear over-determined system of two Eqs.(\ref{sfc}). It follows then that only 
a subclass of solutions to ME is covered by the antiself-dual field strengths (\ref{Gstreng}) 
in agreement with the concept of ``induced nonlinearity''.  On account of the 
theorem of section 3 {\it bounded singular ``sources'' of these fields will carry 
necessarily quantized electric charge}. 

Indeed, the full version of the considered twistor scheme is invariant under 
the so called ``weak'' or {\it restricted} gauge transformations 
[11,21] of the form 
\beg{twistgauge}
\xi \mapsto \alpha(\xi,\tau)\xi, ~~\tau \mapsto \alpha(\xi,\tau) \tau, 
~~C_\mu \mapsto C_\mu + 2\p_\mu \ln \alpha , 
\enq
in which the gauge parameter $\alpha(\xi(x),\tau(x))$ is allowed to depend 
on the space-time coordinates {\it only implicitly}, i.e. only through the 
components of the transforming spinor $\xi(x)$ itself and/or of its twistor 
counterparts $\tau(x)$ defined via the incidence relation (\ref{inc}). In 
order to simplify the above presentation, the gauge symmetry 
(\ref{twistgauge}) has been broken by the scaling of the principal spinor of 
the form $\xi_0 =1$; for manifestly invariant version of the presented twistor 
construction we refer the reader to our works [13,21,29].

For us 
here it's only important to mark that the restricted nature of gauge invariance 
(\ref{twistgauge}) doesn't violate any consideration used in the proof of the 
charge quantization' theorem in section~3. Therefore, all the conditions of the 
theorem are fulfilled: CSD conditions hold good, gauge invariance is ensured 
and the ``wave function'' $G(x)$ is single-valued everywhere outside of the field 
strengths' singularities. Thus, we can state that every bounded singularity 
of Maxwell fields (\ref{Gstreng}) or (\ref{emsp}) obtained from KFC (\ref{kerr}) 
carries necessarily quantized electric charge. Namely, by comparison of 
(\ref{twistgauge}) with (\ref{gtrans}) we find that the (dimensionless) 
``coupling constant'' in our case is equal to $b=2$ and, consequently, for 
admissible value of charge we get  
\beg{Gcharge}
q = Nq_{min} \equiv \frac{N}{4}, ~~~~N=0,\pm 1, \pm 2, ...
\enq

Certainly, numerical value $q_{min} = 1/4$ of the quantum of charge itself 
is here of no particular importance since the dimensional units are as yet 
ambiguous and its dynamical meaning -- as yet not clear. Nonetheless, we are 
free to choose the units (say, of length and of field strength) in the way to 
ensure $q_{min}$ be equal in dimensional units to the elementary 
electron charge. Thereafter, for 
{\it every bounded singularity of any Maxwell field obtained from the KFC 
the charge will be integer multiple of the elementary one}. We 
shall see in the next section, moreover, that fundamental charged solutions 
(of Coulomb-like type) possess {\it precisely} the elementary charge (for 
which $N=\pm 1$ only), the property being specific, to our knowledge, only 
for this construction (compare, e.g., with the Ran\~ada's Coulomb 
Ansatz [9] with the charge {\it arbitrary multiple} of the minimal one).

Finally, let us underline that the above-presented twistor generating 
construction is purely algebraic in origin since one doesn't need to resolve 
here any PDE or even to explicitly integrate in an auxiliary twistor space as 
in the {\it Bateman-Penrose transform} for solutions of 
homogeneous Maxwell~\fnm{j}\fnt{j}{Another simple algebraic procedure 
proposed by I.~Robinson [38] also makes it possible to define a Maxwell field 
for every solution of the KFC. However, this field is null and charge-free} 
and other linear equations [30,27]. Indeed, starting from an 
{\it arbitrary} 
complex function $\Pi(G,\tau^0,\tau^1)$ we differentiate it with respect 
to $\tau^0,\tau^1$ and to $G$ (the latter being the {\it total} 
derivative) and resolve thereafter the algebraical KFC (\ref{kerr}) with 
respect to the only unknown $G$ at any space-time point. Substituting the 
latter into the expression (\ref{Gstreng}) we are able to calculate the 
field strengths at this particular point, i.e. {\it completely locally}.

On the other hand, we can {\it eliminate} the generating field $G$ from the 
system of KFC (\ref{kerr}) and the condition (\ref{sing}) which defines 
the locus of branching points of $G(x)$ together with singular locus of 
field strengths (\ref{emsp}). In the result of elimination we come to one 
(generally complex) constraint of the form 
\beg{motion}
S(u,v,w,\bar w) = S(x,y,z,t) = 0
\enq
which determines, at a fixed moment of time $t$, the shape of  
singular ``source'' of EM field strengths. In ``common case'' one 
complex Eq.(\ref{motion}) corresponds to {\it two} real constraints 
which define a {\it 1-dimensional} ``string-like'' singular object in 
3-dimensional space. As an exception, singularities can be 0-dimensional 
(point charges) or 2-dimensional (membranes). It can be proved [29] that 
for every solution $G(x)$ of KFC or,equivalently, of constraints (\ref{sfc}) 
the function $S(u,v,w, \bar w)$ in Eq.(\ref{motion}) necessarily 
satisfies the complex eikonal equation (see also the last section).  
                         
As the time $t$ in Eq.(\ref{motion}) varies, the latter defines the 
{\it evolution} of the singular locus, i.e. can be regarded as the 
{\it equation of motion} of an extended singular (in particular bounded, 
``particle-like'') object. Examples of shape and 
time evolution of these objects will be presented in the next section.  

In fact, differential constraints (\ref{sfc}) define a well known 
fundamental geometrical structure, namely a {\it shear-free (null 
geodesic) congruence} (SFC) in Minkowski space (see, e.g., [27]). In 
particular, a SFC is formed from the rays of EM wave radiated by an 
arbitrary moving electric charge. By this, the component $G(x)$ of the 
basic projective 2-spinor $\xi^T=\{1,G(x)\}$ 
defines the principal null 4-vector $k_\mu (x)$ tangent to the lines 
of the congruence,
\beg{tang}
k_\mu = \xi^+\sigma_\mu \xi,  
\enq
whereas the famous {\it Kerr theorem} [28,27] asserts that {\it every SFC in 
Minkowsky space can be obtained from KFC (\ref{kerr}) using a generating 
twistor function} $\Pi(G,\tau^0,\tau^1)$.

Moreover, a wide and physically important class of Riemannian metrics 
does exist, of the {\it Kerr-Schild type}
\beg{schild}
g_{\mu\nu} = \eta_{\mu\nu} + H(x)k_\mu k_\nu ,
\enq
for which the property of a congruence to be null geodesic and shear-free 
is preserved. On 
the other hand, a scalar function $H(x)$ can be often defined in the way 
that the metric (\ref{schild}) would satisfy the vacuum or electrovacuum 
Einstein - Maxwell equations [28,20]. Irrespectively to this, the  
{\it curvature singularities} for metrics (\ref{schild}) are fixed via the 
same condition (\ref{sing}) as those of EM fields (\ref{emsp}) and form 
thus, in the framework of our construction, a singular object --  
{\it a unique source of EM and effective gravitational fields}. Note 
that through the correspondence of SFC with asymptotically flat 
solutions of Einstein-Maxwell equations such characteristics as 
(gravitational) mass and angular moment (spin) can be prescribed to 
these singular particle-like objects (see the first two examples from 
the next section).

Finally, potentials of a complex non-Abelian gauge field can be 
defined by the same complex Maxwell 4-potentials $C_\mu(x)$ [11,21]. 
For every generating function $G(x)$ subject to KFC 
{\it these potentials identically satisfy the nonlinear equation of 
the Yang-Mills type!} [11,21].

Thus, the KFC (\ref{kerr}) generates a distinguished subclass of 
interrelated solutions to a large number of (both linear and nonlinear)  
relativistic field equations. These and only these 
solutions are thought of as physically meaningful in the framework 
of the above-presented twistor construction and with respect to the 
concept of induced nonlinearity. 

\section{Singular structure of Maxwell fields obtained via Kerr functional 
condition. Examples}

\noindent
{\bf 1. Quantized Coulomb solution.} Let us start from the previous Ansatz 
(\ref{coulomb}) for superpotential $G(x), \Box G=0$ 
which results in the Coulomb solution (\ref{cfs}) to ME. It's easy to 
check that it satisfies also the eikonal equation (\ref{eik}) but, generally,  
doesn't  satisfy the two differential constraints (\ref{sfc}). Therefore, it 
can't be obtained via twistor algebraic construction based on the KFC (\ref{kerr}). 
The constraints (\ref{sfc}) are fulfilled  only if the numerical factor in 
Eq.(\ref{coulomb}) $q=1/4$, i.e. only for 
{\it elementary} electric charge, in correspondence with the quantization theorem. 

Indeed, only for this value of charge the Ansatz (\ref{coulomb}) can be obtained 
from KFC (\ref{kerr}) via the following generating function:
\beg{coulomb1}
\Pi = G\tau^0 - \tau^1 = G(wG+u)-(vG+\bar w) = wG^2 +2zG - \bar w = 0,
\enq
where $z = (u-v)/2$ and the time parameter $t=(u+v)/2$ doesn't enter the 
defining KFC (\ref{coulomb1}) which turns to be {\it quadratic} in this case. 
Explicitly resolving the latter, we obtain  {\it two branches} of the  
superpotential function 
\beg{stereotwo}
G=\frac{\bar w}{z \pm r}, ~~~r = \sqrt{x^2+y^2+z^2} 
\enq
which, geometrically, correspond to {\it stereographic projection} 
$S^2\mapsto C$ from the North and the South poles respectively  
and, on the other hand, result in the antiself-dual ($\vec {\cal E} = 
i\vec {\cal H}$) EM fields (\ref{Gstreng}) of the form [10,11]
\beg{Gcoul}
-F_{00}={\cal E}_+ = \frac{\bar w}{4r^3}, ~~F_{01}=F_{10}={\cal E}_3 = 
\frac{z}{4r^3}, ~~F_{11} = {\cal E}_- = \frac{w}{4r^3} .
\enq
The field coincides in form with the standard Coulomb one (\ref{cfs}) but 
carries electric charge $q=\pm 1/4$ necessarily equal in modulus to the 
elementary. Notice that the same field can be obtained algebraically from  
expression (\ref{emsp}) if one calculates the derivatives
\beg{twistder}
\Pi_0=G,~~\Pi_1 = -1, ~~\Pi_{00}=\Pi_{01}=\Pi_{11}=0 ~~~~etc.
\enq  

On the other hand, the function $G$ gives rise to a static and  
spherically symmetric SFC (\ref{tang}) and, consequently, -- to a 
Riemannian metric (\ref{schild}) which (for correspondent choice of 
the ``gravitational potential'' $H(x)$ and together with the 
Coulomb field (\ref{Gcoul}) which preserves its form under the 
change of geometry) satisfies the Einstein - Maxwell equations and 
appears to be just the {\it Reissner-Nordstr\"om} solution [28].

\noindent
{\bf 2. Appel-Kerr solution.}  Now let us modify generating function 
(\ref{coulomb1}) as follows:
\beg{Cshift} 
\Pi = G\tau^0-\tau^1 +2iaG = wG^2 + 2(z+ia)G -\bar w = 0,
\enq
$a$ being a real constant. Function (\ref{Cshift}) can be 
obtained from the previous one (\ref{coulomb}) by means of the 
{\it complex translation} $z \mapsto z^* = z + ia$. Solving the 
quadratic equation, we again obtain two branches of the field $G$ 
which have the same form as in (\ref{stereotwo}) but with substitution 
$z \mapsto z^*, r \mapsto r^*$. The branching locus 
of the field $G$ evidently corresponds to the (now complex) condition
\beg{kerrsing}
r^* = \sqrt{x^2+y^2+(z+ia)^2} = 0,
\enq
which defines a {\it ring} of radius equal to $\vert a\vert$, 
\beg{kering}
z=0, ~~~ x^2 + y^2 = a^2 .
\enq
Just on this ring EM field strengths (which correspond again to 
expression (\ref{Gcoul}) with substitution of the ``asterisk''-variables) 
become singular. On the other hand, EM field becomes now globally {\it 
two-fold} changing its sign when one goes round the singular ring. 
However, the ring itself is well-defined in shape and, moreover, 
can be taken as a model of electron, see below.

For this solution the allowed value for electric charge remains, of course, the 
same and equal to the elementary one $q=\pm 1/4$. However, the structure 
of fields becomes much more complicated, and the real and imaginary parts 
can be separated only asymptotically, at distances $r >> |a|$ where 
for the real-part fields ($\Re$-fields) $({\bf E}, {\bf H})$ we get
approximately (in spherical coordinates) [7]
\beg{elec}
E_r \simeq \frac{q}{r^2} (1 -\frac{3 a^2}{2 r^2}(3\cos^2{\theta}-1)), ~~~~
 E_\theta \simeq -\frac{q a^2}{r^4} 3\cos{\theta}\sin{\theta},  
\enq

\beg{magne}
H_r \simeq \frac{2 q a}{r^3}\cos{\theta}, ~~~~ H_\theta \simeq \frac{q a}{ r^3}
\sin{\theta}
\enq
where $q=\pm 1/4$. In view of Eqs. (\ref{elec},\ref{magne}) and (\ref{kerrsing}) 
the 
$\Re$-fields describe a singular ``source'' with elementary electric charge $q$,    
{\it dipole magnetic moment\/} $\mu = qa$ and {\it quadrupole electric moment\/} 
$\vartheta = 2qa^2$, i.e. only those characteristics of elementary particles 
which are observed in experiment.  As to the $\Im$-fields, they are exact 
{\it dual duplicates} of the $\Re$-ones and, according to the arguments of section 3, 
wouldn't, perhaps, manifest themselves in interactions. 

Electric part of the considered solution, i.e. complexification of 
Coulomb field has been obtained (in regard to the historical comments in
[31]) by Appel back in 1887 (see also [20]). Exact Ansatz 
(\ref{Gcoul}) (with substitution of ``asterisk'' - variables) 
turns out to coincide with EM fields for the {\it Kerr-Newman solution} of 
Einstein-Maxwell electrovacuum system. Just this effective metric of 
the type (\ref{schild}) arises from the SFC (\ref{tang}) defined by the 
Appel superpotential $G=\bar w /(z^* \pm r^*)$.

B. Carter demonstrated in [32] that the singular ring of Kerr-Newman solution  
possesses gyromagnetic ratio {\it exactly equal to the Dirac        
value} $g=2$ and proposed a model of electron on this base.
This molel has been thereafter studied in various aspects in the works of 
E.T. Newman [33], A.Ya. Burinskii [34,36], C.A. L\'opez [35] et al. 
In [26] (see also [36]) it has been proved   
that {\it ``particle-like'' (i.e. with bounded singular locus) {\it static}  
solutions which can be obtained from KFC (\ref{kerr}) are exhausted by the 
Appel-Kerr solution of the type (\ref{stereotwo})}. For more detailed 
discussion of these issues we also refer the reader to our papers [7,8].  
                                                                    
\noindent
{\bf Bisingular solution and its modifications.} Solutions of this 
type, in particular with a torus-like singular locus and a nontrivial 
evolution with bifurcations, have been obtained 
and discussed in [13,21]. 

\noindent
{\bf Wave-like singular solutions.} It's easy to see, say, from the 
complex form of ME (\ref{maxC}) that {\it every} null field ${\bf P}_3 =0,
~{\bf P}_- =0,~{\bf P}_+ =P_+(w,u)$ arbitrarily dependent on a pair of spinor 
coordinates $w=x-iy,~u=t+z$, satisfies ME identically. These solutions 
evidently describe EM waves propagating with fundamental velocity along the 
$Z$-direction and somehow distributed in the transversal plane. 
In the framework of the considered twistor construction only a subclass of 
such wave-like solutions can be realized, in the case the 
generating function is independent on one of the twistor variables ($\tau^1$) 
and has the form 
\beg{wavesol}
\Pi=\Pi(G,\tau^0)=\Pi(G,wG+u)=0 .
\enq
In the result, the superpotential $G$ should obey only the first of 
the two constraints (\ref{sfc}) which binds together its dependence on the 
two coordinates and inhibits, in particular, the existence of {\it plane} 
EM waves in the scheme. Instead, from Eq.(\ref{wavesol}) a wide class of 
{\it singular} wave-like solutions can be obtained. In [8] an example 
of the latters has been presented, with a {\it helix-like} singularity
infinite in $Z$- and localized in the transversal direction. More 
interesting is the ``particle-like'' case, with completely bounded 
(localized) singular locus of field strengths, which can be obtained, 
in particular, from a rather simple generating function of the form
\beg{photon}
\Pi=G^2(\tau^0)^2 + a^2G^2-b^2 = 0,
\enq
$a,b=const\in R$ and which will be discussed elsewhere. Such {\it photon-like}  
solutions seem quite unusual and, certainly, need careful study and 
physical interpretation.

\noindent
{\bf ``Cocoon''-like solution.} We conclude with a presentation  
of a time-dependent, axially symmetrical solution for which 
(as well as for the solution generated from Eq.(\ref{photon})) the 
function $G(x)$ can't be resolved from 
the KFC in explicit form. For this, we take the twistor function 
of the form
\beg{polin}
\Pi = G^2 (\tau^0)^2 + (\tau^1)^2 - b^2 G^2 = 0,
\enq
$b>0$ being a real constant. Separating the coordinate $\bar w$
through the substitution 
\beg{subst}
G=\frac{\bar w}{\rho} Y(\rho,u,v), ~~~\rho=\sqrt{x^2+y^2},
\enq
we reduce Eq.(\ref{polin}) to the {\it quartic equation} in $Y$ of the 
form
\beg{axisim}
\rho^2 Y^4 + 2\rho u Y^3 + (u^2+v^2-b^2) Y^2 + 2\rho v Y + \rho^2 = 0, 
\enq
which at initial moment of time $t=(u+v)/2 =0$ can be factorized into a pair 
of quadratic equations
\beg{factor}
(\rho Y^2 + a Y - \rho)(\rho Y^2 + c Y - \rho) = 0,
\enq
with $a,c = z\pm\sqrt{b^2-z^2-2\rho^2}$ respectively. After this,  
analytical analysis of the field $Y$ and of its branching points becomes 
possible. It shows that at $t=0$ for every of the four modes the associated 
EM field at spacial infinity is again Coulomb-like, with elementary charge $q=\pm1/4$, 
whereas the correspondent point singularity 
is located at the axis $\rho=0$, either at $z=+b/\sqrt{2}$ or at $z=-b/\sqrt{2}$. 
Besides, for each mode the fields {\it become singular on the ellipsoidal 
``cocoon'' $z^2+2\rho^2 =b^2$ covering the point singularity}. More 
detailed analysis of this field and its time evolution will be presented 
elsewhere.               
  
\section{The algebrodynamical approach and its biquaternionic realization} 

There exists a great number of links between the twistor 
construction based on KFC (\ref{kerr}) 
and the general approach to field theory developed in our works 
during the last two decades -- the {\it algebrodynamics}. As we have seen, the KFC is 
deeply related, apart from the twistors themselves, to a lot of peculiar 
algebraical and geometrical 
structures, namely to the shear-free congruences  
and effective Riemannian metrics, to the 
vector fields covariantly constant with respect to exeptional affine {\it 
connection of the Weyl-Cartan type} [10,11,21] and, in algebrodynamical 
theory, -- to the Cauchy-Riemann 
type equations generalized to the noncommutative algebra of complex quaternions. 
All these structures constitute in essence a unique fundamental entity 
which can 
be described in diverse yet equivalent languages. According to the philosophy 
of algebrodynamics [10,37], such an abstract structure can be considered 
as a candidate for the primodial {\it Code of Nature}, and all the genuine 
physical laws should directly 
follow from its intrinsic properties which can only be carefully examined. 

In our works [10,11,15,7,21] we have taken the algebra of complex quaternions 
(or {\it biquaternions}) $B$ as the fundamental 
structure which completely determines both physical geometry and dynamics.  
Physical fields ($B$-fields) were assumed to be just the {\it differentiable 
functions of $B$-variable}. 
Owing to the {\it noncommutativity} of $B$-algebra, the differentiability conditions,  
i.e. the {\it generalized Cauchy-Riemann equations} (GCRE) turned to be nonlinear and 
over-determined. Besides, they were found to be naturally 
Lorentz and gauge invariant, and have been taken as the only fundamental field 
equations of a unified algebraic field theory. 

Remarkably, the nonlinear Lorentz invariant {\it complex eikonal equation} (CEE)
\beg{eik}
(\p_t S)^2 - (\p_x S)^2 - (\p_y S)^2 - (\p_z S)^2 = 0 
\enq
should be satisfied for every (spinor) component $S(t,x,y,z)\in C$ of any 
differentiable function of B-variable (B-field) [10,11] 
and is of great importance for the theory, similar to that of the 
linear Laplace equation for complex analysis. 

Aside from this, in [29] the {\it general solution of the CEE}  
(\ref{eik}) has been obtained (through the analysis of its intrinsic twistor 
structure) which turned out to consist of {\it two different classes}. For 
the first one every solution is implicitly defined just by the KFC (\ref{kerr}),
whereas solutions of the second class can be obtained from them via 
the elimination procedure described in section 5 and constitute the 
{\it singular locus} (\ref{motion}) for the eikonal fields of the first class. 
For them, the primary GCRE take the following invariant form [10,11,8]:
\beg{GCRE}
d\xi = C dX \xi ,
\enq
where the 2-spinor $\xi^T(x)=\{1, G(x)\}$ and the $2\times 2$ complex matrix 
$C(x)$ correspond to the superpotential and complex 4-potentials 
respectively which were introduced in sections 4,5 in context of generating 
constructions of solutions to ME. Conditions of integrability of  
GCRE (\ref{GCRE}) result, as before, in CSD conditions for potentials 
$C(x)$ and, consequently, -- in satisfaction of ME, WE as well as 
Yang-Mills type equations ``on shell''. 

On the other hand, the potentials $C(x)$ can be eliminated from 
over-determined  GCRE (\ref{GCRE}), and the spinor function $G(x)$ 
turns therafter to satisfy the differential constraints (\ref{sfc}) and 
to define, therefore, a shear-free congruence. Then, with respect to 
the Kerr theorem (section 5), {\it every solution $\{\xi(x),C(x)\}$ of 
GCRE (\ref{GCRE}) can be algebraically obtained via KFC (\ref{kerr}) from 
a generating twistor function $\Pi(G,\tau^0,\tau^1)$}.

Thus, the presented version of the algebrodynaimcs reduce completely to the analysis 
of the KFC which has been partially accomlished in sections 5,6. Apart from 
Maxwell fields, many other {\it dynamical} field structures can be naturally 
broght into correspondence with the KFC (see the end of section 5). The 
singular loci of these fields coincide in shape and in dynamics being defined 
from the ``caustic condition'' (\ref{sing}). In the case the singular locus is 
spatially bounded, its connected components can be considered as a set of 
particle-like objects with self-consistent and topologically nontrivial 
shape and evolution. These objects manifest some properties of 
elementary particles, carrying, in particular, discrete and integer multiple 
electric charge and the magnetic moment and spin with gyromagnetic ratio 
which is inherent just to fermions.  

We can say, in conclusion, that in the algebrodynamical approach physical 
problems reduce to those of algebraic geometry and of theory of singularities of 
differentiable mappings. By this, no extraneous assumptions are 
allowed (for, say, better correspondence with physical phenomenology). 
Future study will show to which extent the 
real properties of elementary particles are encoded in those of  
fundamental mathematical structures dealt with in the paper.

\section{Acknowledgments}
The author are greatly indebted to V.~I.~Zharikov and to V.~N.~Trishin.

\nonumsection{References}

\end{document}